\documentstyle[12pt,epsf]{article}
\begin{document}
\begin{center}
\begin{Large}
{\bf  Search for New Particles Decaying to Dijets\\
in $p\bar{p}$ Collisions at $\sqrt{s}=1.8$ TeV }
\end{Large}
\end{center}
\font\eightit=cmti8
\def\r#1{\ignorespaces $^{#1}$}
\hfilneg
\begin{sloppypar}
\noindent
F.~Abe,\r {13} M.~G.~Albrow,\r 7 S.~R.~Amendolia,\r {23} D.~Amidei,\r {16}
J.~Antos,\r {28} C.~Anway-Wiese,\r 4 G.~Apollinari,\r {26} H.~Areti,\r 7
M.~Atac,\r 7 P.~Auchincloss,\r {25}
% ...
\end{sloppypar}

\vskip .025in
\begin{center}
(CDF Collaboration)
\end{center}

\vskip .025in
\begin{center}
\r 1  {\eightit Argonne National Laboratory, Argonne, Illinois 60439} \\
\r 2  {\eightit Istituto Nazionale di Fisica Nucleare, University of Bologna,
I-40126 Bologna, Italy} \\
\r 3  {\eightit Brandeis University, Waltham, Massachusetts 02254} \\
\r 4  {\eightit University of California at Los Angeles, Los
Angeles, California  90024} \\
\r 5  {\eightit University of Chicago, Chicago, Illinois 60637} \\
\r 6  {\eightit Duke University, Durham, North Carolina  27708} \\
\r 7  {\eightit Fermi National Accelerator Laboratory, Batavia, Illinois
60510} \\
\r 8  {\eightit Laboratori Nazionali di Frascati, Istituto Nazionale di Fisica
               Nucleare, I-00044 Frascati, Italy} \\
\r 9  {\eightit Harvard University, Cambridge, Massachusetts 02138} \\
\r {10} {\eightit University of Illinois, Urbana, Illinois 61801} \\
\r {11} {\eightit Institute of Particle Physics, McGill University, Montreal
H3A 2T8, and University of Toronto,\\ Toronto M5S 1A7, Canada} \\
\r {12} {\eightit The Johns Hopkins University, Baltimore, Maryland 21218} \\
\r {13} {\eightit National Laboratory for High Energy Physics (KEK), Tsukuba,
Ibaraki 305, Japan} \\
\r {14} {\eightit Lawrence Berkeley Laboratory, Berkeley, California 94720} \\
\r {15} {\eightit Massachusetts Institute of Technology, Cambridge,
Massachusetts  02139} \\
\r {16} {\eightit University of Michigan, Ann Arbor, Michigan 48109} \\
\r {17} {\eightit Michigan State University, East Lansing, Michigan  48824} \\
\r {18} {\eightit University of New Mexico, Albuquerque, New Mexico 87131} \\
\r {19} {\eightit Osaka City University, Osaka 588, Japan} \\
\r {20} {\eightit Universita di Padova, Instituto Nazionale di Fisica
          Nucleare, Sezione di Padova, I-35131 Padova, Italy} \\
\r {21} {\eightit University of Pennsylvania, Philadelphia,
        Pennsylvania 19104} \\
\r {22} {\eightit University of Pittsburgh, Pittsburgh, Pennsylvania 15260} \\
\r {23} {\eightit Istituto Nazionale di Fisica Nucleare, University and Scuola
               Normale Superiore of Pisa, I-56100 Pisa, Italy} \\
\r {24} {\eightit Purdue University, West Lafayette, Indiana 47907} \\
\r {25} {\eightit University of Rochester, Rochester, New York 14627} \\
\r {26} {\eightit Rockefeller University, New York, New York 10021} \\
\r {27} {\eightit Rutgers University, Piscataway, New Jersey 08854} \\
\r {28} {\eightit Academia Sinica, Taiwan 11529, Republic of China} \\
\r {29} {\eightit Superconducting Super Collider Laboratory, Dallas,
Texas 75237} \\
\r {30} {\eightit Texas A\&M University, College Station, Texas 77843} \\
\r {31} {\eightit Texas Tech University, Lubbock, Texas 79409} \\
\r {32} {\eightit University of Tsukuba, Tsukuba, Ibaraki 305, Japan} \\
\r {33} {\eightit Tufts University, Medford, Massachusetts 02155} \\
\r {34} {\eightit University of Wisconsin, Madison, Wisconsin 53706} \\
\r {35} {\eightit Yale University, New Haven, Connecticut 06511} \\
\r a    {\eightit Visitor} \\
\end{center}

\renewcommand{\baselinestretch}{2}
\large
\normalsize

\begin{center}
{\bf Abstract}
\end{center}
We have used $19$ pb$^{-1}$ of data collected with the Collider Detector at
Fermilab to search for new particles decaying to dijets.
We exclude at 95\% confidence level models containing the following new
particles:
axigluons with mass between 200 and 870 GeV/c$^2$,
excited quarks with mass between 80 and 570 GeV/c$^2$, and color octet
technirhos with mass between 320 and 480 GeV/c$^2$.

\vspace*{0.1in}
\vspace*{1.0in}

Within the framework of perturbative QCD two-jet events are expected to arise
in proton-antiproton collisions from hard parton-parton scattering.  The
outgoing scattered partons manifest themselves as hadronic jets. The predicted
two-jet mass spectrum falls rapidly with increasing two-jet mass.  Many
extensions of the standard model predict the existence of new massive objects
that couple to quarks and gluons, and result in resonant structures in the
two-jet mass spectrum.  In this paper we report a search for narrow
resonances in the two-jet mass spectrum measured in proton-antiproton
collisions at a center of mass energy $\sqrt{s}=1.8$ TeV.

 In addition to this general search, we specifically search for the
following six resonance phenomena.
First, in a model where the symmetry group $SU(3)$ of QCD is replaced by the
chiral symmetry $SU(3)_L\times SU(3)_R$, there are axial vector particles
called
axigluons A~\cite{ref_axi}.  The axigluon is produced and decays in the
quark-antiquark channel ($A\rightarrow q \bar{q}$); here we have assumed they
decay only to the quarks in the standard model.
Second, if quarks are composite
particles then excited states $q^*$ are expected~\cite{ref_qstar}; we search
for mass degenerate excited quarks in the quark-gluon channel
($q^* \rightarrow qg$).
Third, models of walking technicolor~\cite{ref_trho}, which seek to explain
electro-weak symmetry breaking via the dynamics of a new interaction among
techniquarks, predict the presence of color octet technirhos $\rho_T$.  We
consider such a model in which the $\rho_T$ are mass degenerate and decay to
dijets only ($\rho_T \rightarrow g \rightarrow q\bar{q}$ or $gg$); in this
model the technipion is too massive to be a decay product of the $\rho_T$.
Fourth and fifth,
models which propose new gauge symmetries often predict new
gauge bosons~\cite{ref_gauge} which decay to quarks ($W^\prime$,$Z^\prime
\rightarrow
q\bar{q}$). Here we assume standard model couplings and when calculating the
cross section include a K-factor~\cite{ref_K_factor} to account for higher
order terms. Finally,
superstring theory suggests that $E_6$ may be the grand unified strong and
electro-weak gauge group; $E_6$ models predict the presence of scalar diquarks
$D$ and $D^c$~\cite{ref_diquark} which decay to quarks ($D\rightarrow
\bar{u}\bar{d}$ and $D^c \rightarrow ud$).  We assume electromagnetic strength
couplings and mass degenerate diquarks.

A detailed description of the Collider Detector at Fermilab (CDF) can be found
elsewhere~\cite{ref_CDF}.
We use a coordinate system with $z$ along the proton
beam, transverse coordinate perpendicular to the beam, azimuthal angle $\phi$,
polar angle $\theta$, and pseudorapidity $\eta=-\ln \tan(\theta/2)$.
Jets are reconstructed as localized energy depositions in the CDF calorimeters
which are constructed in a tower geometry.
The jet energy E and momentum
$\vec{P}$ are defined as the scalar and vector sums respectively of calorimeter
tower energies inside a cone of radius
$R=\sqrt{(\Delta\eta)^2 + (\Delta\phi)^2}=0.7$
centered on the jet direction. E and $\vec{P}$ are corrected
for calorimeter non-linearities, energy lost in uninstrumented regions and
outside the clustering cone, and
energy gained from the underlying event.
The jet energy corrections increase the jet energies on average by roughly
27\%(15\%) for 50 GeV (500 GeV) jets.  Full details of jet reconstruction and
jet energy corrections at CDF can be found elsewhere~\cite{ref_jet}.

We define the dijet system as the
two jets with the highest transverse momentum in the event (leading jets)
and define the dijet mass
$m=\sqrt{(E_1 + E_2)^2 - (\vec{P}_1 + \vec{P}_2)^2}$.  The dijet mass
resolution is approximately 10\%.
Our data sample was obtained in the 1992-93 running period using four single
jet triggers with thresholds on the uncorrected cluster transverse energies
of 20, 50, 70 and 100 GeV.  After jet energy corrections these trigger samples
were used to measure the dijet mass spectrum above 150, 241, 265, and 353
GeV/c$^2$ respectively.  At these mass thresholds the trigger efficiencies for
the
four triggers were 1.0, 0.99, 0.87, and 0.89, and the four data samples
corresponded to integrated luminosities of 0.038, 0.66, 3.2, and 19.1
pb$^{-1}$ after prescaling.  Offline we required that
both jets have pseudorapidity $|\eta|<2$ and a scattering angle in the dijet
center of mass frame
$|\cos\theta^*| = |\tanh[(\eta_1-\eta_2)/2]| < 2/3$. The $\cos\theta^*$ cut
provides uniform acceptance as a function of mass and reduces the QCD
background which  peaks at $|\cos\theta^*|=1$.
To maintain the projective nature of the calorimeter towers, the $z$ position
of the event vertex was required to be within 60 cm of the center of the
detector; this cut was 94\% efficient. Backgrounds from cosmic-ray
interactions were rejected if the energy deposited in the central hadronic
calorimeters occurred at times other than the the $p\bar{p}$ crossing.
Remaining backgrounds from cosmic-rays, beam halo, and detector noise produced
events with unusually large or unbalanced energy depositions and were removed
by requiring \mbox{${\not\!\!E_T}/\sqrt{\sum E_T}<6$} and $\sum E_T< 2$ TeV,
where \mbox{${\not\!\!E_T}$} is the missing transverse
energy~\cite{ref_missing_et} and $\sum E_T$ is the total transverse energy in
the event.

\begin{figure}[tbh]
\hspace*{-0.5in}
\vspace*{-1.2in}
\epsffile[36 81 540 650]{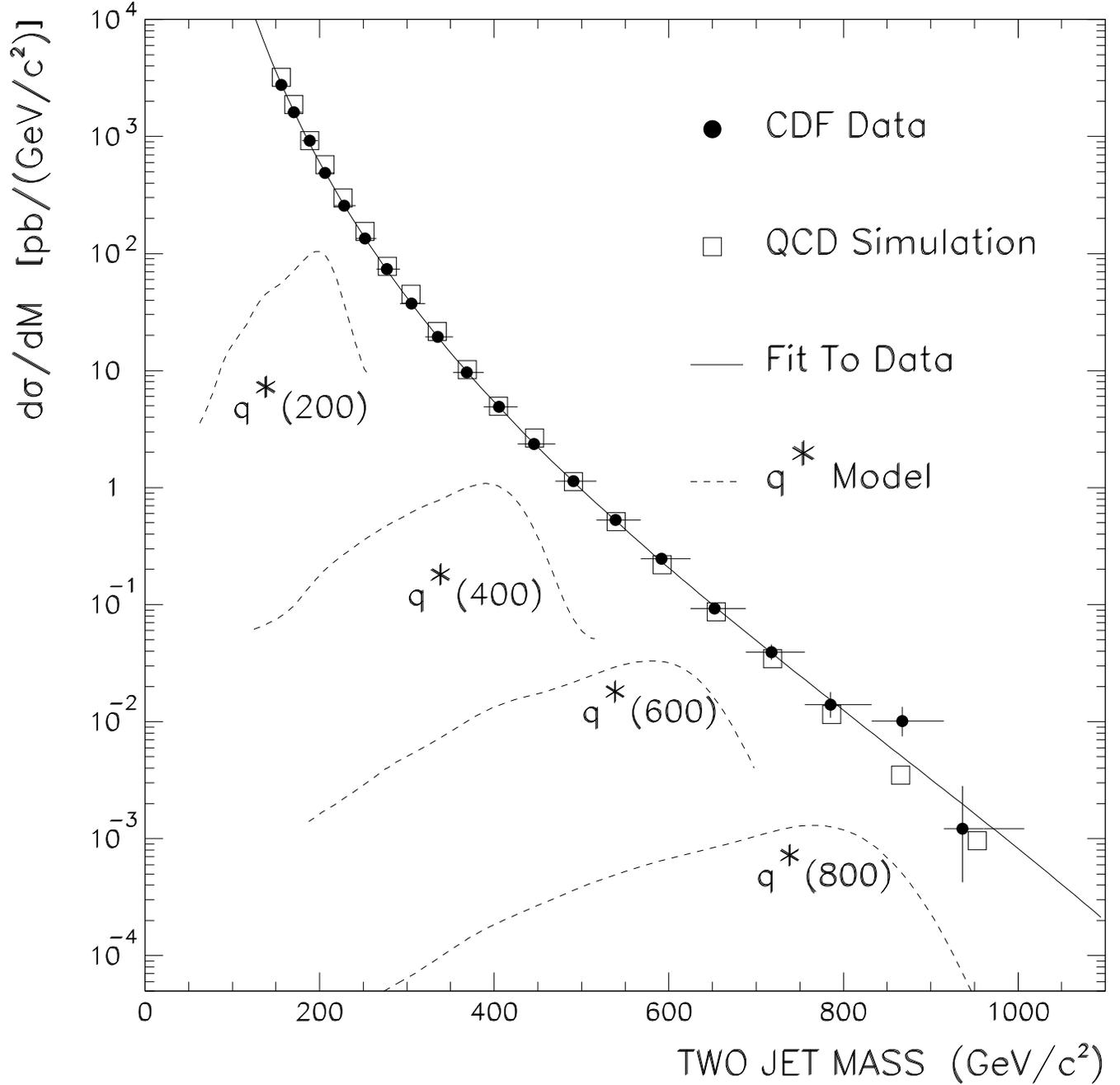}
\caption[Dijet Mass, Background and Excited Quark Signal]{
The dijet mass distribution (circles) compared to a QCD simulation (boxes) and
fit to a smooth parameterization (solid curve). Also shown are simulations
of excited quark signals in the
CDF detector (dashed curves). In the data and simulations we require that both
jets have pseudorapidity $|\eta|<2.0$ and the dijet satisfies
$|\cos\theta^*|<2/3$.}
\label{fig_mass_qstar}
\end{figure}

\begin{figure}[tbh]
\hspace*{-0.5in}
\vspace*{-1.0in}
\epsffile[36 81 540 650]{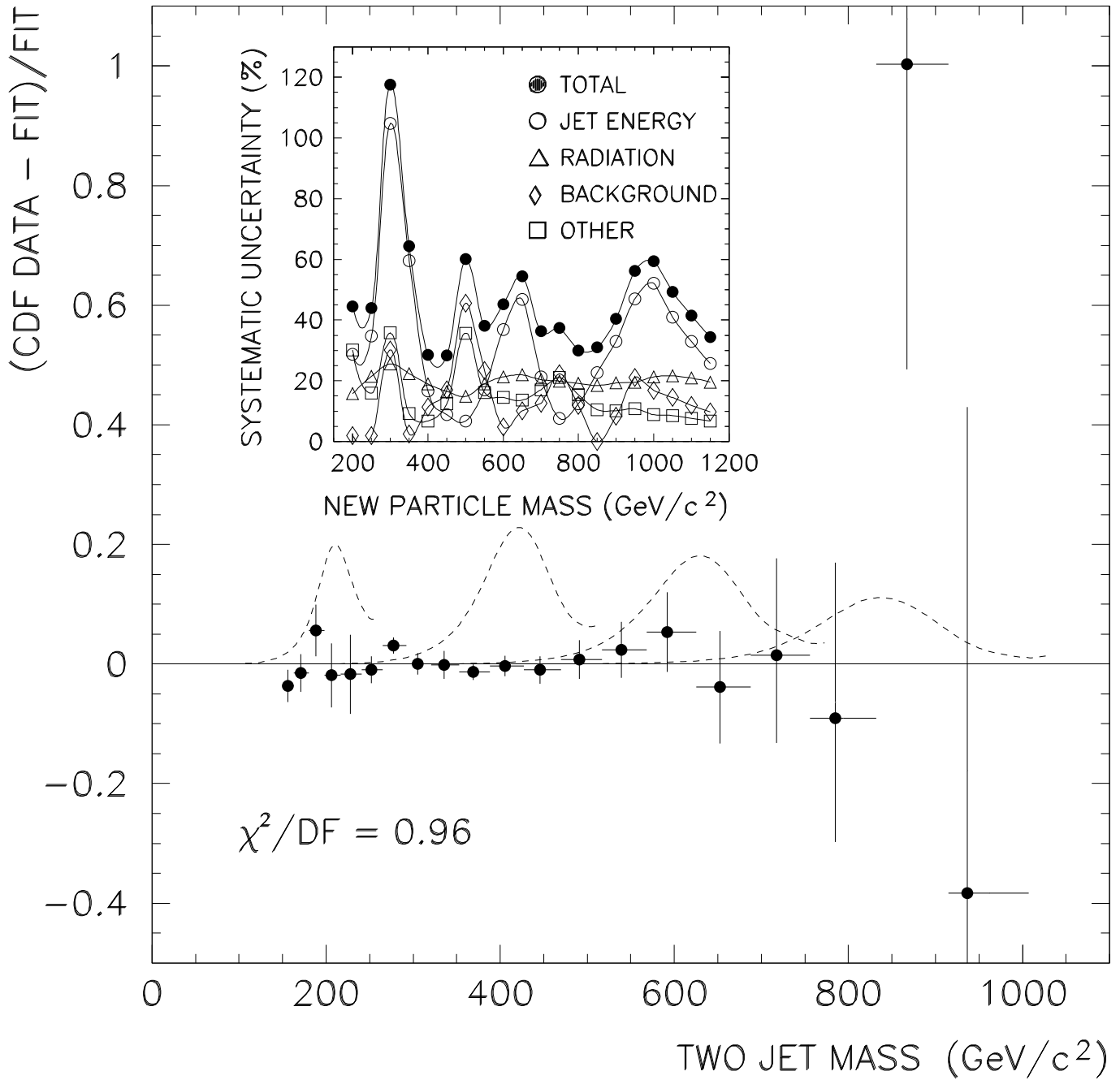}
\caption[Dijet Mass and Background Fit on a Linear Scale]{
The fractional difference between the dijet mass distribution (points) and
a smooth background fit (solid line) is compared to simulations of
excited quark signals in the CDF detector (dashed curves). The inset shows the
systematic uncertainty for a new particle signal (see text).}
\label{fig_mass_lin}
\end{figure}

In Fig.~\ref{fig_mass_qstar} we present the inclusive dijet mass
distribution for $p\bar{p}\rightarrow$ 2 leading jets + X, where X can be
anything including additional jets.
The dijet mass
distribution has been corrected for trigger and $z$ vertex inefficiencies.
We plot the differential cross section versus the mean dijet mass in
bins equal to the dijet mass resolution (RMS$\sim 10$\%). The data is compared
to a QCD prediction~\cite{ref_pythia} which includes a simulation of the CDF
detector. The QCD prediction uses CTEQ2L parton distributions~\cite{ref_CTEQ},
a renormalization scale $\mu=P_T$, and is normalized to the data.
We also fit the data with the parameterization
$d\sigma/dm = A(1-m/\sqrt{s})^N/m^P$ with parameters  A, N and P.
This parameterization gives a good description of both the observed
distribution ($\chi^2/DF=.96$) and the QCD prediction ($\chi^2/DF=.75$).
The fit to the observed distribution gave $A=(6.4\pm 0.1)\times 10^{15}$
pb/(GeV/c$^2$), $N=5.512\pm0.002$, and $P=6.69\pm 0.09$, where the quoted
errors
 are statistical only.
Fig.~\ref{fig_mass_qstar} shows the background fit on a
logarithmic scale, and
Fig.~\ref{fig_mass_lin} shows the fractional difference between the data and
background fit on a linear scale.  The fit shows no significant evidence for
any new particle. Upward fluctuations appearing
in the data near 250, 550 and 850 GeV/c$^2$ have statistical significance 2.3,
1.3 and 1.8 standard deviations when interpreted as ``signals''
for new particles at these masses.
However, this minimal significance is reduced by roughly
a factor of 2 after incorporating systematic uncertainties (discussed later).

To set limits on dijet resonances it is sufficient to determine the mass
resolution for only one new particle type assuming each new particle's
natural half-width ($\Gamma/2$) is small compared to the dijet mass
resolution. This is the case for these models of axigluons
($\Gamma/2M\approx 0.05$), excited quarks
($\Gamma/2M\approx 0.02$), color octet technirhos ($\Gamma/2M\approx 0.01$),
new
gauge bosons ($\Gamma/2M\approx 0.01$), and $E_6$ diquarks ($\Gamma/2M\approx
0.004$ for $D$ and $0.001$ for $D^c$). In Figs.~\ref{fig_mass_qstar} and
\ref{fig_mass_lin} we show the
predicted mass resolution for excited quarks (q*) using the
PYTHIA Monte Carlo~\cite{ref_pythia} and a CDF detector
simulation. The mass resolution has a Gaussian core (RMS/M $\sim 0.1$) from
jet energy resolution and a long tail towards low mass from QCD radiation.
We have used the q* mass resonance curves in Figs.~\ref{fig_mass_qstar} and
\ref{fig_mass_lin} to
model the shape of all new particles decaying to dijets.
As in our previous search for excited quarks~\cite{ref_qstar_prl},
we perform a binned maximum likelihood fit
of the data to both the background parameterization and the signal
hypothesis. The method gave a Poisson likelihood as a function of the signal
cross section.  This was done independently
at 20 different values of new particle mass from 200 to 1150 GeV/c$^2$,
resulting in 20 statistical likelihood distributions.

\begin{figure}[tbh]
\hspace*{-0.5in}
\vspace*{-1.3in}
\epsffile[36 81 540 650]{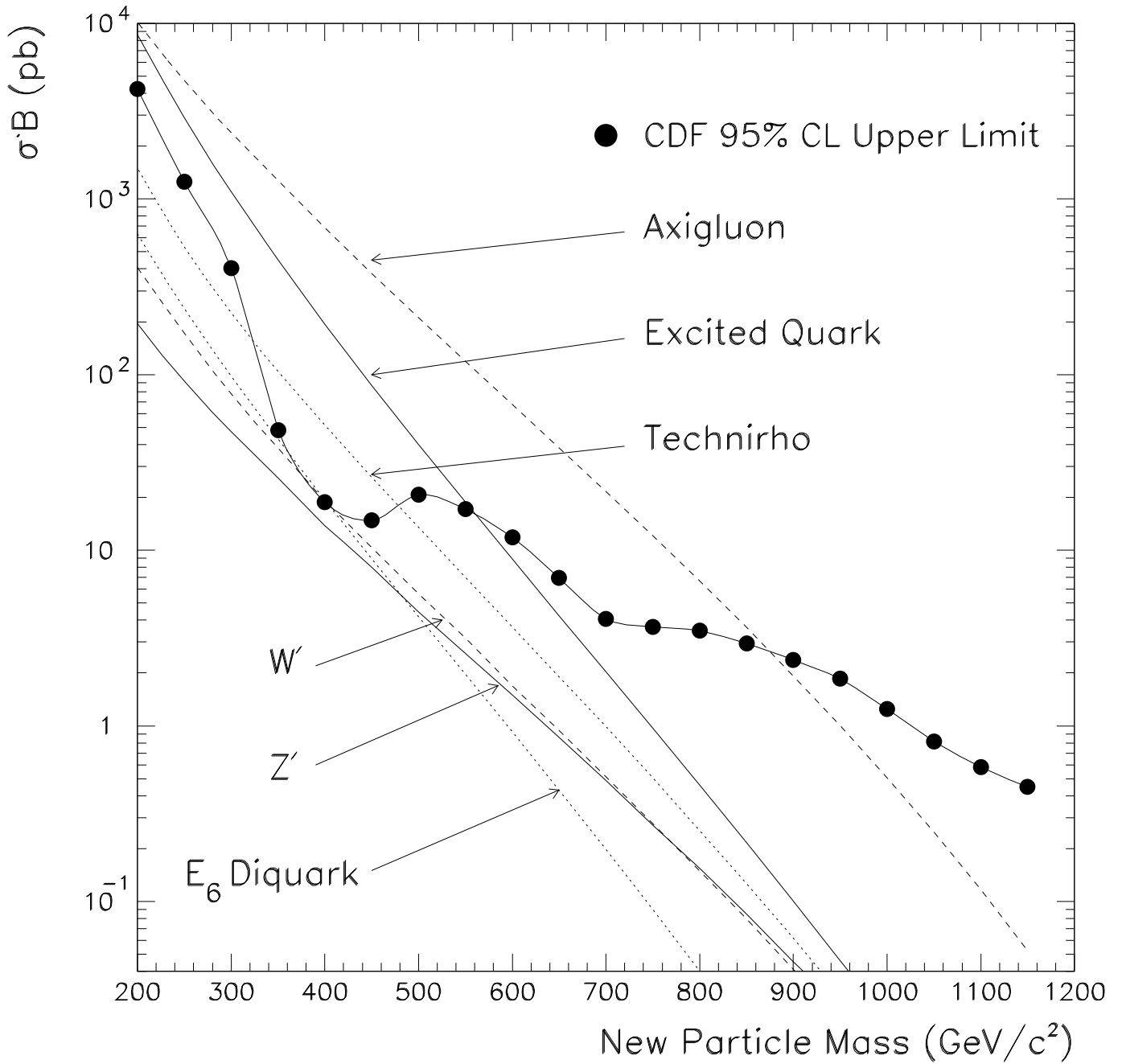}
\caption[Upper Limit on Cross Section for New Particles]{
The upper limit on the cross section times branching
ratio for new particles decaying to dijets (points) is compared to
theoretical predictions for axigluons~\cite{ref_axi}, excited
quarks~\cite{ref_qstar}, color octet technirhos~\cite{ref_trho}, new
gauge bosons $W^{\prime}$ and $Z^{\prime}$~\cite{ref_gauge,ref_K_factor}, and
$E_6$ diquarks~\cite{ref_diquark}. The limit and theory curves require that
both
jets have pseudorapidity $|\eta|<2.0$ and the dijet satisfies
$|\cos\theta^*|<2/3$.}
\label{fig_limit}
\end{figure}

Systematic uncertainties on the cross section for observing a new particle
in the CDF detector
are shown in Fig.~\ref{fig_mass_lin}.
Each systematic uncertainty on the fitted signal cross section was determined
by varying the source of uncertainty by $\pm 1\sigma$ and refitting.
In decreasing order of importance the sources
of uncertainty are the 5\% jet energy scale
uncertainty, QCD radiation's effect on the mass resonance line
shape, the background parameterization, trigger efficiency, jet energy
resolution, jet energy scale of CDF calorimeters relative to the central
calorimeter, luminosity and efficiency.
For example, at 300 GeV/c$^2$ reducing the jet energy by 5\% centers the
resonance on an upward fluctuation, and increases the fitted signal by
over 100\%.
The total systematic uncertainty was found by adding the above
sources in quadrature. We convoluted each of the 20
likelihood distributions with the corresponding total Gaussian systematic
uncertainty, and found the 95\% confidence level
(CL) upper limit shown in Fig.~\ref{fig_limit}.

\begin{figure}[tbh]
\hspace*{-0.1in}
\vspace*{-0.2in}
\epsffile[72 144 550 650]{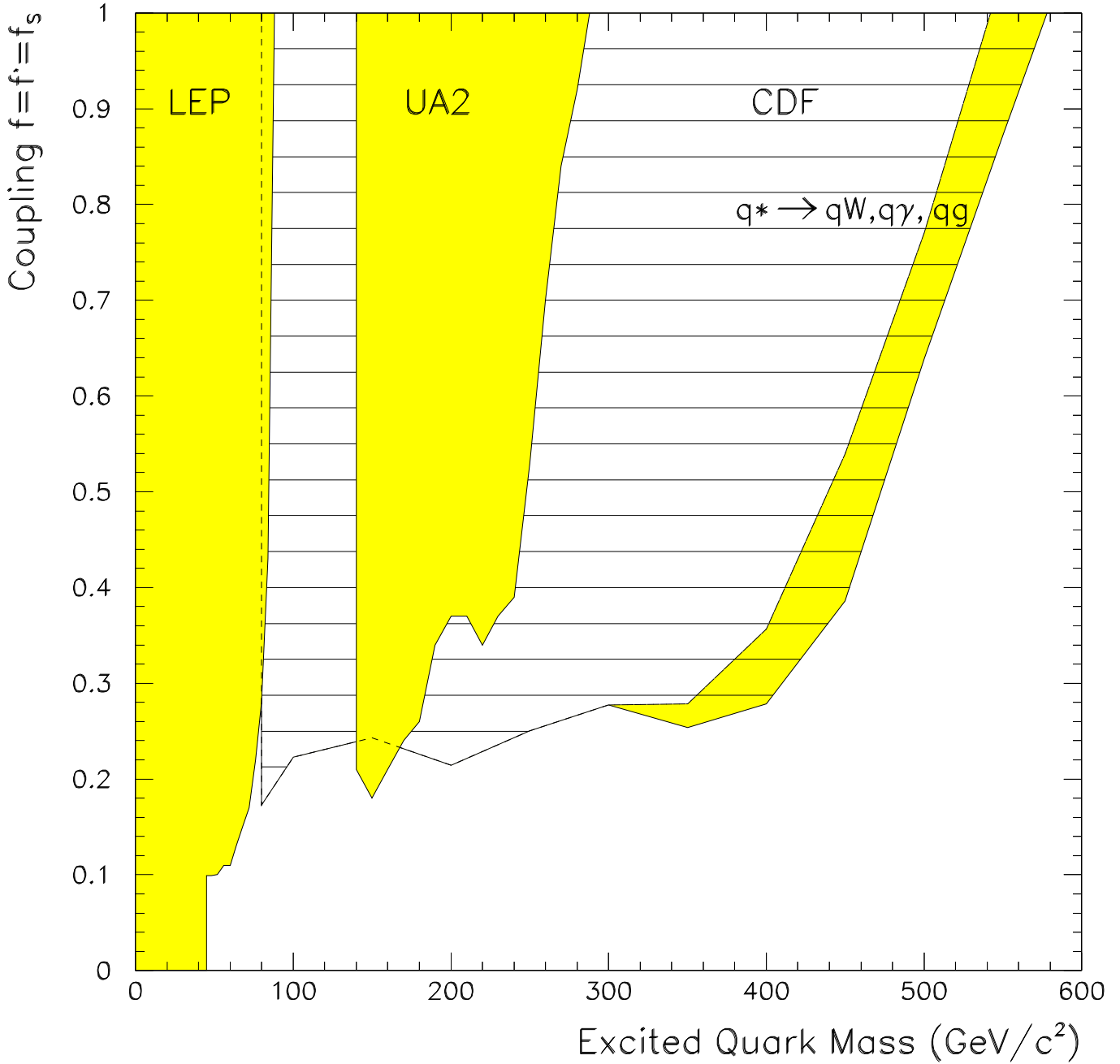}
\caption[Limit on Excited Quarks in Mass vs. Coupling]{
The region of the coupling vs. mass plane excluded by previous CDF
measurements~\cite{ref_qstar_prl} in the $q^* \rightarrow q\gamma$ and
$q^* \rightarrow qW$ channels (hatched region) is extended by combining them
with this search in the $q^* \rightarrow qg$ channel (shaded and hatched
region). The CDF excluded regions are compared to the regions excluded by
LEP and UA2 (shaded regions)~\cite{ref_other}.}
\label{fig_coupling}
\end{figure}

In Fig.~\ref{fig_limit} and Table I we compare our measured upper limit
on the cross section times branching ratio for a new particle decaying to
dijets to the theoretical predictions.
The predictions are lowest order with one-loop strong coupling $\alpha_s(m^2)$
and CTEQ2L parton
distributions~\cite{ref_CTEQ}. Branching fractions to top quarks are included
but do not add to the dijet mass resonance cross section. New particle decay
angular distributions are included, and we  required
$|\eta|<2$ and $|\cos\theta^*|<2/3$. We exclude at 95\% CL new particles in
mass regions for
which the theory curve lies above our upper limit. For axigluons~\cite{ref_axi}
we exclude the region $200<M_A<870$ GeV/c$^2$, significantly extending the
previous CDF exclusions of $120<M_A<210$ GeV/c$^2$~\cite{ref_CDFaxi1}
and $240<M_A<640$ GeV/c$^2$~\cite{ref_CDFaxi2} and the
UA1 exclusion of $110<M_A<310$ GeV/c$^2$~\cite{ref_UA1axi}.
For the first time we exclude a model of technicolor~\cite{ref_trho}
with color octet technirhos in the mass range $320<M_{\rho_T}<480$ GeV/c$^2$.
{}From Fig.~\ref{fig_limit} we exclude excited quarks in the
mass range $200<M^*<560$ GeV/c$^2$ at 95\% CL. This limit from dijets
can be improved by combining it with published limits in the
$\gamma$+jet and the W+jet channel~\cite{ref_qstar_prl} (by multiplying the
likelihood distributions). Combining all three channels excludes excited quarks
in the mass interval $80<M^*<570$ GeV/c$^2$ for standard model couplings.
The excluded regions in the coupling~\cite{ref_qstar} vs. mass plane are shown
in
Fig.~\ref{fig_coupling} compared to previous excluded regions.  The cross
section for new gauge bosons and $E_6$ diquarks is too small to be excluded by
our data.

In conclusion, the measured dijet mass spectrum is a smoothly falling
distribution within statistics.  We see no significant evidence for new
particle production and set limits on axigluons, excited
quarks, and color octet technirhos.

We thank the Fermilab staff and the technical staffs of the participating
institutions for their vital contributions. We also thank Ken Lane for
providing software we used to calculate the technirho cross section. This work
was
supported by the U.S. Department of Energy and National Science Foundation;
the Italian Istituto Nazionale di Fisica Nucleare; the Ministry of Science,
Culture, and Education of Japan; the Natural Sciences and Engineering Research
Council of Canada; the A. P. Sloan Foundation; and the Alexander von
Humboldt-Stiftung.

\clearpage

\renewcommand{\baselinestretch}{1.4}
\large
\normalsize

\begin{table}[tbh]
\vspace*{-0.5in}
\hspace*{-0.5in}
\begin{tabular}{|l|l|l|l|l|l|l|l|}\hline
 & 95\% CL  & \multicolumn{6}{c|}{Theory Cross Section $\times$ Branching
Ratio}\\  \cline{3-8}
Mass & $\sigma$.B Limit & A & $q^*$ & $\rho_T$ &
$W^\prime$ & $Z^\prime$ & $D+D^c$\\
 {\footnotesize GeV/c$^2$} & (pb) & (pb) & (pb) & (pb) & (pb) & (pb) & (pb)\\
\hline
\ 200&$4.2\times 10^3$&$1.0\times 10^4$&$8.6\times 10^3$& $1.5\times 10^3$
&                       $4.1\times 10^2$&$1.9\times 10^2$& $6.3\times 10^2$\\
\ 250&$1.3\times 10^3$&$4.7\times 10^3$&$2.9\times 10^3$ & $5.3\times 10^2$
&                       $1.7\times 10^2$&$9.2\times 10^1$&$2.4\times 10^2$\\
\ 300&$4.0\times 10^2$&$2.4\times 10^3$&$1.1\times 10^3$ & $2.2\times 10^2$
&                       $7.8\times 10^1$&$4.7\times 10^1$&$9.8\times 10^1$\\
\ 350&$4.8\times 10^1$&$1.3\times 10^3$&$4.5\times 10^2$ & $1.0\times 10^2$
&                       $3.8\times 10^1$&$2.6\times 10^1$&$4.3\times 10^1$\\
\ 400&$1.9\times 10^1$&$6.8\times 10^2$&$1.9\times 10^2$ & $5.2\times 10^1$
&                       $1.9\times 10^1$&$1.4\times 10^1$&$1.9\times 10^1$\\
\ 450&$1.5\times 10^1$&$3.8\times 10^2$&$8.7\times 10^1$ & $2.7\times 10^1$
&                       $1.0\times 10^1$&$8.0$            &$8.9$           \\
\ 500&$2.1\times 10^1$&$2.1\times 10^2$&$4.0\times 10^1$ & $1.4\times 10^1$
&                       $5.6$&           $4.4$            &$4.2$           \\
\ 550&$1.7\times 10^1$&$1.2\times 10^2$&$1.9\times 10^1$ & $7.3$
&                       $3.2$&           $2.6$            &$2.0$           \\
\ 600&$1.2\times 10^1$&$6.8\times 10^1$&$8.9$            & $3.8$
&                       $1.7$&           $1.5$            &$9.2\times
10^{-1}$\\
\ 650&7.0             &$3.8\times 10^1$&$4.2$            & $2.0$
&                  $9.4\times 10^{-1}$&$8.6\times 10^{-1}$&$4.3\times
10^{-1}$\\
\ 700&4.1             &$2.2\times 10^1$& 2.0             & 1.1
&                  $5.1\times 10^{-1}$&$4.9\times 10^{-1}$&$2.0\times
10^{-1}$\\
\ 750&3.7        &$1.2\times 10^1$&$9.7\times 10^{-1}$& $5.5\times 10^{-1}$
&                  $2.8\times 10^{-1}$&$2.7\times 10^{-1}$&$9.0\times
10^{-2}$\\
\ 800&3.5             & 6.7            & $4.6\times 10^{-1}$&$2.8\times
10^{-1}$
&                  $1.5\times 10^{-1}$&$1.5\times 10^{-1}$&$4.0\times
10^{-2}$\\
\ 850&2.9             & 3.6            & $2.2\times 10^{-1}$&$1.4\times
10^{-1}$
&                  $7.9\times 10^{-2}$&$8.5\times 10^{-2}$&$1.7\times
10^{-2}$\\
\ 900&2.4             & 1.9            & $1.0\times 10^{-1}$&$6.9\times
10^{-2}$
&                  $4.1\times 10^{-2}$&$4.6\times 10^{-2}$&$7.3\times
10^{-3}$\\
\ 950&1.9             & 1.0            & $4.6\times 10^{-2}$&$3.3\times
10^{-2}$
&                  $2.1\times 10^{-2}$&$2.4\times 10^{-2}$&$3.0\times
10^{-3}$\\
1000&1.2            &$5.1\times 10^{-1}$&$2.1\times 10^{-2}$&$1.6\times
10^{-2}$
& -- & -- & -- \\
1050&$8.1\times 10^{-1}$&$2.4\times 10^{-1}$& $9.4\times 10^{-3}$ & --
& -- & -- & -- \\
1100&$5.8\times 10^{-1}$&$1.2\times 10^{-1}$& $4.2\times 10^{-3}$ & --
& -- & -- & -- \\
1150&$4.5\times 10^{-1}$&$5.2\times 10^{-2}$& $1.8\times 10^{-3}$ & --
& -- & -- & -- \\
\hline
\end{tabular}
\label{tab_fit}
\end{table}

\renewcommand{\baselinestretch}{2}
\large
\normalsize
\vspace*{-0.3in}
Table I: As a function of new particle mass we list our 95\% CL upper limit on
cross section times branching ratio and the theoretical prediction for
axigluons (A), excited quarks ($q^*$), color octet technirhos ($\rho_T$),
new gauge bosons ($W^\prime$ and $Z^\prime$), and $E_6$ diquarks ($D+D^c$).
The limit and predictions require that both
jets have pseudorapidity $|\eta|<2.0$ and the dijet satisfies
$|\cos\theta^*|<2/3$.
\clearpage

\clearpage

\clearpage

\clearpage

\end{document}